\documentclass[11pt,twoside]{atmp}

\usepackage{amsmath,amssymb}

\usepackage[all]{xy}
\usepackage{color}

\usepackage{epsf}
\usepackage{subeqnarray}

\DeclareMathOperator{\e}{e} 

\def\r{\;\;\;}

\def\fns{\footnotesize}
\def\scz{\scriptsize}
\def\nrs{\normalsize}

\def\dl{\displaylimits}
\def\ds{\displaystyle}
\def\si{\mbox{\scz i}}
\def\i{\mbox{i}}
\def\be{\begin{equation}}
\def\ee{\end{equation}}
\def\bse{\begin{subeqnarray}}
\def\ese{\end{subeqnarray}}
\def\ba{\begin{array}}
\def\ea{\end{array}}
\def\beq{\begin{eqnarray}}
\def\eeq{\end{eqnarray}}
\def\bmp{\begin{minipage}}
\def\emp{\end{minipage}}
\def\nn{\nonumber}
\def\np{\newpage}

\def\bfl{\begin{flushleft}}
\def\efl{\end{flushleft}}
\def\bfr{\begin{flushright}}
\def\efr{\end{flushright}}
\def\bc{\begin{center}}
\def\ec{\end{center}}
\def\ben{\begin{enumerate}}
\def\een{\end{enumerate}}
\def\bit{\begin{itemize}}
\def\eit{\end{itemize}}

\newcommand{\vx}{\vspace*{1.0mm}}
\newcommand{\vy}{\vspace*{2.5mm}}

\newcommand{\vn}{\vspace*{-1.5mm}}
\newcommand{\vm}{\vspace*{-2.5mm}}

\newcommand{\vl}{\vspace*{-5.0mm}}

\newcommand{\hy}{\hspace*{2.5mm}}

\begin{document}

\title[Optical properties of nano-crystals]
{Exact Microtheoretical Approach to Calculation of Optical
Properties of Ultralow Dimensional Crystals}
%
%
\author[J.P.\v{S}etraj\v{c}i\'{c}]{Jovan P. \v{S}etraj\v{c}i\'{c}}
\address{Department of Physics, Faculty of Sciences, University
of Novi Sad, \\ Trg Dositeja Obradovi\'{c}a 4, 21.000 Novi Sad,
Vojvodina $-$ Serbia}
\address{and}
\address{Academy of Sciences and Arts of the Republika Srpska, \\
Trg srpskih vladara 2/II, 78.000 Banja Luka, Republic of Srpska $-$
B\&H}
\addressemail{jovan.setrajcic@df.uns.ac.rs}
\begin{abstract}
The main problem in theoretical analysis of structures with strong
confinement is the fact that standard mathematical tools: differential
equations and Fourier's transformations are no longer applicable. In this
paper we have demonstrated that method of Green's functions can be
successfully used on low-dimension crystal samples, as a consequence
of quantum size effects. We can illustrate modified model through the
prime cubic structure molecular crystal: bulk and ultrathin film.
Our analysis starts with standard exciton Hamiltonian with definition
of commutative Green's function and equation of motion.
We have presented detailed procedure of calculations of
Green's functions, and further dispersion law, distribution of states
and relative permittivity for bulk samples. After this, we have followed the
same procedures for obtaining the properties of excitons in ultra-thin films.
The results have been presented graphically.
Besides modified method of Green's functions we have shown that the
exciton energy spectrum is discrete in film structures (with
number of energy levels equal to the number of atomic planes of the
film). Compared to the bulk structures, with continual absorption zone, in
film structures exist resonant absorption peaks. With increased film
thickness differences between bulk and film vanish.
\end{abstract}



\maketitle

\section{Introduction}

Interest in exciton subsystem studies appeared due to the fact that
excitons are responsible for dielectric, optical (absorption,
dispersion of light, luminescence), photoelectric and other
properties of crystals  [1$-$3]. Studies of excitons in
crystalline subsystems culminated with laser invention.

In recent years theoretical investigations of
quasi-two-dimensional exciton subsystems (nanostructures) were
intensified, especially in the field of thin films, not only to
obtain fundamental information regarding dielectric
properties of these materials but also because of their wide
practical use (nanoelectronics, optoelectronic [4$-$6],
light energy conversion [9,10]...). What is unique for these
structures is that they have changed properties
compared to their bulk analogues [6$-$8].

We studied the basic physical characteristics of ultrathin
dielectrics $-$ molecular crystalline nanofilms [11,12], which
could be used as surface layers for protection of
electronic components or as special light filters.

This paper analyzes the influence of border-film structure on
the energy spectrum of excitons (exciton dispersion law). Special
attention was paid to the presence and spatial distribution of localized
exciton states. Optical properties of these dielectric films were
also investigated (their dielectric permittivity was determined).
Results obtained in this work were compared with the similar results
for the case of ideal infinite crystals, in order to find most
important differences between these two systems.

These analyzes may be conducted using methods of two-time,
temperature-dependant Green's functions that are often
used in quantum theory of solid state [13$-$15]. With adequately
incorporated statistics, this method is being successfully applied
in calculations of microscopic and macroscopic, as well
as balanced and non-balanced properties of crystals.

A question which justifiably arises is related to the mean of calculating Green's functions, which
are "borrowed" from the quantum field theory and whose
definition, i.e. usage, is based on variables with continuous spectra
in unlimited (both direct and impulse) space!

This work proves that the method of Green's functions may be
successfully applied onto crystalline samples of such small dimensions
that the quantum size effects are relevant [16]. In order to
illustrate adaptation of this method, we will observe molecule
crystal with a simple cubic structure: spatially unlimited (bulk)
and strongly limited along one axis (ultra thin film). Our intention
is to show a technique of application of Green's function
onto spatially limited systems, so we excluded from
calculation all really existing parameters: more complex crystalline
structure, changes in boundary film parameters etc.

\section{Excitons in bulk-structures}

The discussion of dielectric properties of an ideal (with no defects, vacancies, etc)
unlimited molecular crystal will start with standard exciton Hamiltonian, which has
following form in configuration space [1,15,17]: \vx
\begin{equation}
    H = H_{0} + \sum_{\vec{n}} \Delta_{\vec{n}} \, P_{\vec{n}}^{+}
    P_{\vec{n}} + \sum_{\vec{n}, \vec{m}} X_{\vec{n} \vec{m}} \,
    P_{\vec{n}}^{+} P_{\vec{m}} +
    \sum_{\vec{n}, \vec{m}} Y_{\vec{n} \vec{m}} \, P_{\vec{n}}^{+}
    P_{\vec{n}} P_{\vec{m}}^{+} P_{\vec{m}} \,,
\end{equation}
where $P_{\vec{n}}^{+}$ and $P_{\vec{n}}$ represent creation and
annihilation exciton operators at the node (site) $\vec{n}$ of the
crystalline lattice. Quantity $\Delta_{ \vec{n}}$ represents the
energy of the exciton localized at the $ \vec{n}$ node, while the
quantities $X_{\vec{n}, \vec{m}}$ and $Y_{\vec{n}, \vec{m}}$ represent
matrix elements of the exciton transfer from the node $\vec{n}$ to
the node $\vec{m}$.

Properties of the model exciton system may be analyzed using
commutation Paulian Green's function [13,14,18]: %
\begin{equation}
    \Gamma_{ \vec{n} \vec{m}} (t) \equiv \langle \langle P_{\vec{n}}(t)
    \mid P_{\vec{m}}^{+}(0) \rangle \rangle =
    \Theta (t) \, \langle \left[ P_{\vec{n}} (t), \,
    P_{\vec{m}}^{+} (0) \right] \rangle \,,
\end{equation}
which satisfies following equation of motion:
\begin{flalign}
    \i \hbar \, \frac{d}{d t} \, \Gamma_{ \vec{n} \vec{m}} (t) &=
    \i \hbar \, \delta (t) \, \langle \left[ P_{\vec{n}} (t), \,
    P_{\vec{m}}^{+} (0) \right] \rangle + \\
    &+ \Theta (t) \langle \left[ P_{\vec{n}}(t), H \right] \,
    P_{\vec{m}}^{+} (0) - P_{\vec{m}}^{+}(0) \left[ P_{\vec{n}}(t),
    H \right] \rangle \nn \,.
\end{flalign}
Using commutation relations for Pauli operators [18,19]:
\begin{equation} \ba{c}
    \ds \left[ P_{\vec{n}}, P_{\vec{m}}^{+} \right] = \left(1-2
    P_{\vec{n}}^{+} P_{\vec{n}} \right) \, \delta_{\vec{n} \vec{m}}
    \,; \vx \\
    \ds \left[ P_{\vec{n}}, P_{\vec{m}} \right] = \left[
    P_{\vec{n}}, P_{\vec{m}}^{+} \right] = 0 \,; \r
    P_{\vec{n}}^{2} = \left(P_{\vec{n}}^{+} \right)^{2} = 0 \,, \ea
\end{equation}
we obtain the equation of motion for Paulian Green's function:
\begin{flalign}
    \i \hbar \, \frac{d}{d t} \, \Gamma_{ \vec{n} \vec{m}} (t) &=
    \i \hbar \, \delta (t) \delta_{\vec{n} \vec{m}} \left(1 - 2
    \langle P_{\vec{n}}^{+} P_{\vec{n}} \rangle \right) + \nn \\
    &+ \Delta_{\vec{n}} \, \Gamma_{\vec{n} \vec{m}} (t)
    + \sum_{\vec{l}} X_{\vec{n} \vec{l}} \, \Gamma_{\vec{l}
    \vec{m}} (t) - \\
    &- 2 \sum_{\vec{l}} X_{\vec{n} \vec{l}} \, \mathcal{T}_{\vec{n}
    \vec{n} \vec{l} \vec{m}} (t) + 2 \sum_{\vec{l}} Y_{\vec{n}
    \vec{l}} \, \mathcal{T}_{\vec{l} \vec{l} \vec{n} \vec{m}}(t) \nn
\end{flalign}
expressed by $\mathcal{T}_{\vec{n} \vec{n} \vec{l} \vec{m}} (t) =
\langle \langle P_{\vec{n}}^{+}(t) P_{\vec{n}}(t) P_{\vec{l}}(t)
\mid P_{\vec{m}}^{+}(0) \rangle \rangle -$ Paulian Green's function
of the higher (third) order.

The basic problem with exciton theory is the fact that
Pauli-operators $P^{+}$ and $P$ are not Bose or Fermi operators,
but a certain hybrid of both with a kinematics described by expression (4),
that is Fermian for one mode and Bosonian for different modes. For precise
analysis of exciton systems, which encompass effects of
inter-exciton interaction, simple replacing of Pauli-operators with
Bose-operators is not enough. Therefore, in Hamiltonian (1),
Pauli-operators are replaced by their exact Bosonian represents
[20]:
\begin{flalign}
    &P = \left[ \sum_{\nu = 0}^{\infty} \frac{(-2)^{\nu}}{(1+
    \nu)!} \left( B^{+} \right)^{\nu} B^{\nu} \right]^{\frac{1}{2}}
    B \,; \nn \\
    &P^{+} = B^{+} \left[ \sum_{\nu = 0}^{\infty}
    \frac{(-2)^{\nu}}{(1+\nu)!} \left( B^{+} \right)^{\nu} B^{\nu}
    \right]^{\frac{1}{2}} \,; \\
    &P^{+} P = \sum_{\nu = 0}^{\infty} \frac{(-2)^{\nu}}{(1+ \nu)!}
    \left( B^{+} \right)^{\nu + 1} B^{\nu + 1} \,. \nn
\end{flalign}
Our goal is to adapt Green's function method to spatially
quantum (discrete, not continuous) structures and to
see the influence of spatial limits and disturbances of inner
translational symmetry on changes of their macroscopic physical
properties. Paulian Green's functions from equation (7) will be
therefore expressed using appropriate Bosonian Green's functions
on the basis of approximate expressions following from (6):
\begin{equation} \ba{c} \ds
    P \approx B - B^{+} B B \,; \r
    P^{+} \approx B^{+} - B^{+} B^{+} B \,; \vx \\
    \ds P^{+} P \approx B^{+}B - B^{+} B^{+} B B \,. \ea
\end{equation}
By this we obtain:
\begin{flalign}
    &\Gamma_{\vec{n} \vec{m}}(t) = \langle \langle P_{\vec{n}}(t)
    \mid P_{\vec{m}}^{+} (0) \rangle \rangle = \langle \langle
    B_{\vec{n}}(t) \mid B_{\vec{m}}^{+} (0) \rangle \rangle - \nn \\
    &- \langle \langle B_{\vec{n}} (t) \mid B_{\vec{m}}^{+}(0)
    B_{\vec{m}}^{+} (0) B_{\vec{m}}(0) \rangle \rangle -
    \langle \langle B_{\vec{n}}^{+} (t) B_{\vec{n}}(t)
    B_{\vec{n}}(t) \mid B_{\vec{m}}^{+}(0) \rangle \rangle + \\
    &+ \langle \langle B_{\vec{n}}^{+} (t) B_{\vec{n}}(t)
    B_{\vec{n}}(t) \mid B_{\vec{m}}^{+}(0) B_{\vec{m}}^{+} (0)
    B_{\vec{m}}(0) \rangle \rangle \nn \,.
\end{flalign}
Further, by decoupling higher Green's functions using known Bose-commutation
relations:
\begin{equation}
    \left[ B_{\vec{k}}, B_{\vec{l}}^{+} \right] = \delta_{\vec{k}
    \vec{l}} \,; \r \left[B_{\vec{k}}, B_{\vec{l}} \right] =
    \left[B_{\vec{k}}^{+}, B_{\vec{l}}^{+} \right] = 0
\end{equation}
and by introducing retarded (Bosonian) Green's function:
\begin{equation}
    \langle \langle B_{\vec{n}}(t) \mid B_{\vec{m}}^{+} (0)
    \rangle \rangle = G_{\vec{n} \vec{m}} (t) \,,
\end{equation}
terms in expression (8) become:
\begin{flalign}
    \langle \langle B_{\vec{n}} (t) \mid B_{\vec{m}}^{+}(0)
    B_{\vec{m}}^{+} (0) B_{\vec{m}}(0) \rangle \rangle
    = \Theta (t) \langle \left[ B_{\vec{n}}, B_{\vec{m}}^{+}
    B_{\vec{m}}^{+} B_{\vec{m}} \right] \rangle &= \nn \\
    = \Theta (t) \left( \langle \left( \delta_{\vec{n}
    \vec{m}} + B_{\vec{m}}^{+} B_{\vec{n}} \right)
    B_{\vec{m}}^{+} B_{\vec{m}} \rangle \right.
    - \left. \langle B_{\vec{m}}^{+} B_{\vec{m}}^{+} B_{\vec{m}}
    B_{\vec{n}} \rangle \right) = 2 G_{\vec{n} \vec{m}} (t)
    \mathcal{N}_{0} \,; & \nn \\
    \langle \langle B_{\vec{n}}^{+} (t) B_{\vec{n}}(t)
    B_{\vec{n}}(t) \mid B_{\vec{m}}^{+}(0) \rangle \rangle = 2
    G_{\vec{n} \vec{m}} (t) \mathcal{N}_{0} \,; & \\
    \r \langle \langle B_{\vec{n}}^{+} (t) B_{\vec{n}}(t)
    B_{\vec{n}}(t) \mid B_{\vec{m}}^{+}(0) B_{\vec{m}}^{+} (0)
    B_{\vec{m}}(0) \rangle \rangle
    = 2 R_{\vec{n} \vec{m}} (t) \, G_{\vec{n} \vec{m}}^{2} (t)
    \,, & \nn
\end{flalign}
where $\mathcal{N}_{0}$ is concentration of excitons, and
$R_{\vec{n} \vec{m}} (t)$ is advanced Green's function:
\begin{equation} \ba{l} \ds
    \mathcal{N}_{0} = \langle B^{+} B \rangle = \frac{1}{N}
    \sum_{\vec{k}} \left( \mbox{ e}^{\hbar \omega_{_{0}}
    (\vec{k})/\theta} - 1 \right)^{-1}; \\
    \ds R_{\vec{n} \vec{m}} (t) = \langle \langle
    B_{\vec{n}}^{+}(t) \mid B_{\vec{m}}(0) \rangle
    \rangle \,. \ea
\end{equation}
When expressions (10) and (11) are substitued into eq.(8) we obtain
final expression for Paulian Green's function expressed using
Bosonian Green's functions:
\begin{equation}
    \Gamma_{\vec{n} \vec{m}}(t) = \left( 1 - 4 \, \mathcal{N}_{0}
    \right) \, G_{\vec{n} \vec{m}} (t) +
    2 R_{\vec{n} \vec{m}} (t) \, G_{\vec{n} \vec{m}}^{2} (t)
    + O (\mathcal{N}^{2}) \,.
\end{equation}
For Paulian Green's functions of higher order ($\mathcal{T}_{\vec{a}
\vec{a} \vec{b} \vec{c}}$) at the left side of Green's function we
simply replace Pauli operators by Bose-operators, and on the right
side approximation (7) takes place. In this way it follows:
\begin{flalign}
    \mathcal{T}_{\vec{a} \vec{a} \vec{b} \vec{c}} &= \langle \langle
    P_{\vec{a}}^{+}(t) P_{\vec{a}}(t) P_{\vec{b}}(t) \mid
    P_{\vec{c}}^{+}(0) \rangle \rangle
    = \langle \langle B_{\vec{a}}^{+}(t) B_{\vec{a}}(t)
    B_{\vec{b}}(t) \mid B_{\vec{c}}^{+}(0) \rangle \rangle - \nn \\
    &- \langle \langle B_{\vec{a}}^{+}(t) B_{\vec{a}}(t)
    B_{\vec{b}}(t) \mid B_{\vec{c}}^{+}(0) B_{\vec{c}}^{+}(0)
    B_{\vec{c}}(0) \rangle \rangle =  \\
    &= \mathcal{N}_{0} G_{\vec{b} \vec{c}}(t) + \mathcal{N}_{\vec{b}
    \vec{a}} G_{\vec{a} \vec{c}}(t)
    - 2 R_{\vec{a} \vec{c}}(t) \, G_{\vec{b} \vec{c}}(t) \,
    G_{\vec{a} \vec{c}}(t) + O(\mathcal{N}_{0}^{2}) \nn \,.
\end{flalign}
Expressions for $\Gamma_{\vec{n} \vec{m}} $, $ \mathcal{T}_{\vec{n}
\vec{n} \vec{l} \vec{m}} $ and $ \mathcal{T}_{\vec{l} \vec{l}
\vec{n} \vec{m}} $, which are expressed using Bosonian Green's
functions, are substituted in a equation of movement for Paulian Green's
function (5):
\begin{flalign}
    \i \hbar \, \frac{d}{d t} \, \left[\left( 1 - 4 \, \mathcal{N}_{0}
    \right) \, G_{\vec{n} \vec{m}} (t) +  2 R_{\vec{n} \vec{m}} (t)
    \, G_{\vec{n} \vec{m}}^{2} (t) \right]
    = \i \hbar \, \delta (t) \delta_{\vec{n} \vec{m}} \left( 1 -
    2 \langle P_{\vec{n}}^{+} P_{\vec{n}} \rangle \right) &+ \nn \\
    + \Delta_{\vec{n}} \left[ \left(1 - 4 \mathcal{N}_{0} \right) \,
    G_{\vec{n} \vec{m}} (t) + 2 R_{\vec{n} \vec{m}}(t) \, G_{\vec{n}
    \vec{m}}^{2}(t) \right] &+ \nn \\
    + \sum_{\vec{l}} X_{\vec{n}\vec{l}} \left[ \left(1 - 4
    \mathcal{N}_{0} \right) \, G_{\vec{l} \vec{m}} (t) + 2 R_{\vec{l}
    \vec{m}}(t) \, G_{\vec{l} \vec{m}}^{2}(t) \right] &- \\
    \r - 2 \sum_{\vec{l}} X_{\vec{n}\vec{l}} \left[\mathcal{N}_{0}
    G_{\vec{l} \vec{m}} (t) + \mathcal{N}_{\vec{l} \vec{n}} G_{\vec{n}
    \vec{m}}(t) \right. - \left. 2 R_{\vec{n} \vec{m}}(t) \,
    G_{\vec{l}\vec{m}}(t) \, G_{\vec{n}\vec{m}} (t) \right] &+ \nn \\
    \r + 2 \sum_{\vec{l}} Y_{\vec{n}\vec{l}} \left[\mathcal{N}_{0}
    G_{\vec{n} \vec{m}} (t) + \mathcal{N}_{\vec{n} \vec{l}} G_{\vec{l}
    \vec{m}}(t) \right. - \left. 2 R_{\vec{l} \vec{m}}(t) \,
    G_{\vec{n}\vec{m}}(t) \, G_{\vec{l}\vec{m}} (t) \right] \, &. \nn
\end{flalign}
Since concentration of Frenkel's excitons in molecular crystals is
very low ($\mathcal{N} < 1$ \%), the equation above may be solved in
the lowest approximation (this approximation is appropriate for neglecting
anharmonic and nonlinear effects, i.e. non-calculating
higher orders terms of exciton-exciton as well as exciton-phonon
interactions, in agreement to estimates in [1$-$3,13$-$15]):
\begin{flalign}
    &\langle P_{\vec{n}}^{+} P_{\vec{n}} \rangle \approx \langle
    B_{\vec{n}}^{+} B_{\vec{n}} \rangle = \mathcal{N}_{0} \approx 0
    \,; \nn \\
    &\mathcal{N}_{\vec{a} \vec{b}} \approx 0 \,; \r G \cdot G \approx
    0 \,; \r G \cdot R \approx 0 \,. \nn
\end{flalign}
"Decoupled" eq.(15) then obtains the following form:
\begin{equation}
    \i \hbar \frac{d}{d t} G_{\vec{n}\vec{m}} (t) = \i \hbar
    \delta (t) \delta_{\vec{n}\vec{m}} + \Delta_{\vec{n}}
    G_{\vec{n}\vec{m}} (t) + \sum_{\vec{l}} X_{\vec{n}\vec{l}}
    G_{\vec{l}\vec{m}} (t) \,.
\end{equation}

It is important to notice that this equation has to
be obtained in the same form starting from effective Bosonian exciton Hamiltonian in
harmonic approximation:
\begin{equation*}
    H_{\mbox{\scz ex}} = \sum_{\vec{n}} \Delta_{\vec{n}} \, B_{\vec{n}}^{+}
    B_{\vec{n}} + \sum_{\vec{n}, \vec{m}} X_{\vec{n} \vec{m}} \,
    B_{\vec{n}}^{+} B_{\vec{m}} \,,
\end{equation*}
and estimating Bosonian Green's function (10):
\begin{equation*}
    G_{ \vec{n} \vec{m}} (t) = \langle \langle B_{\vec{n}}(t)
    \mid B_{\vec{m}}^{+}(0) \rangle \rangle =
    \Theta (t) \, \langle \left[ B_{\vec{n}} (t), \,
    B_{\vec{m}}^{+} (0) \right] \rangle \,,
\end{equation*}
with its equation motion:
\begin{flalign}
    \i \hbar \, \frac{d}{d t} \, G_{ \vec{n} \vec{m}} (t) &=
    \i \hbar \, \delta (t) \, \langle \left[ B_{\vec{n}} (t), \,
    B_{\vec{m}}^{+} (0) \right] \rangle + \nn \\
    &+ \Theta (t) \langle \left[ B_{\vec{n}}(t), H_{\mbox{\scz ex}}
    \right] \, B_{\vec{m}}^{+} (0) - B_{\vec{m}}^{+}(0)
    \left[ B_{\vec{n}}(t), H_{\mbox{\scz ex}} \right] \rangle \nn \,.
\end{flalign}

This (by time) differential equation (16) is solved using temporal
Fourier's transformation:
\begin{equation}
    f_{\vec{a} \vec{b}} (t) = \int\dl_{-\infty}^{+\infty} d \omega
    f_{\vec{a} \vec{b}} (\omega) \, \e^{-\si \omega t} \,; \r
    \delta (t) = \frac{1}{2 \pi} \int\dl_{-\infty}^{+\infty} d \omega
    \, \e^{-\si \omega t} \,,
\end{equation}
and thus we obtain:
\begin{equation}
    \hbar \omega G_{\vec{n}\vec{m}} (\omega) = \frac{\i \hbar}{2
    \pi} \delta_{\vec{n} \vec{m}}+ \Delta_{\vec{n}}
    G_{\vec{n}\vec{m}} (\omega) + \sum_{\vec{l}} X_{\vec{n}\vec{l}}
    G_{\vec{l}\vec{m}} (\omega) \,.
\end{equation}
By using nearest neighbors approximation ($\vec{l} \rightarrow
\vec{n} \pm \vec{\lambda}_{i} $): $\vec{n} \pm \vec{\lambda}_{1} =
n_{x} \pm 1, n_{y}, n_{z}$; \  $\vec{n} \pm \vec{\lambda}_{2} =
n_{x} , n_{y} \pm 1, n_{z}$; \ $\vec{n} \pm \vec{\lambda}_{3} =
n_{x} , n_{y}, n_{z} \pm 1$ and taking into account that we are
observing an ideal cubic structure, where exciton energy is the
same at every node, and the energy transfer between neighbors is also
the same: $\Delta_{\vec{a}} \equiv \Delta$; \ $X_{\vec{a}, \vec{a}
\pm \vec{\lambda}_{i}} \equiv X_{i}$ \ $i \in \{x, y, z \}$,
equation above is taking following form:
\begin{flalign}
    \hbar \omega \, G_{n_{x}n_{y}n_{z}, m_{x}m_{y}m_{z}}
    (\omega) &\equiv \hbar \omega \, G_{n_{x}n_{y}n_{z}, \vec{m}}
    (\omega) = \nn \\
    &= \frac{\i \hbar}{2 \pi} \delta_{n_{x}n_{y}n_{z},
    \vec{m}} + \Delta \, G_{n_{x}n_{y}n_{z},
    \vec{m}} (\omega) + \nn \\
    &+ X_x \left[G_{n_{x}+1, n_{y}n_{z}; \vec{m}}(\omega)
    + G_{n_{x}-1, n_{y}n_{z}; \vec{m}} (\omega) \right] + \\
    &+ X_y \left[ G_{n_{x} n_{y}+1,n_{z}; \vec{m}}(\omega)
    + G_{n_{x} n_{y}-1, n_{z}; \vec{m}}(\omega) \right] +
    \nn \\
    &+ X_z \left[ G_{n_{x} n_{y}n_{z}+1; \vec{m}} (\omega)
    + G_{n_{x} n_{y}n_{z}-1; \vec{m}} (\omega) \right]
    \nn \,.
\end{flalign}
Since the crystal is unlimited, when solving this linear
differential equation we may use the full spatial Fourier's
transformation:
\begin{equation}
    f_{\vec{a}\vec{b}}(\omega) = \frac{1}{N} \sum_{\vec{k}}
    f_{\vec{k}}(\omega) \e^{\si \vec{k} (\vec{a} - \vec{b})} \,;
    \r \delta_{\vec{a}\vec{b}} = \frac{1}{N} \sum_{\vec{k}} \e^{\si
    \vec{k} (\vec{a} - \vec{b})} \,.
\end{equation}
By using these transformations and by composing the equation
above, we obtain:
\begin{flalign}
    \hbar \omega \, G_{\vec{k}}(\omega) &= \frac{\i \hbar}{2 \pi}
    + \Delta \, G_{\vec{k}} (\omega) + \nn \\
    &+ 2 \left( X_{x} \cos a_{x} k_{x} + X_{y} \cos a_{y} k_{y} +
    X_{z} \cos a_{z} k_{z} \right) \, G_{\vec{k}} (\omega) \,,
\end{flalign}
and from there we may express Green's function:
\begin{flalign}
    G_{\vec{k}} (\omega) &= \frac{\i \hbar}{2 \pi} \,
    \left[ \hbar \omega - \Delta - \right. \nn \\
    &- 2 \left. \left( X_{x} \cos a_{x} k_{x} +
    X_{y} \cos a_{y} k_{y} + X_{z} \cos a_{z} k_{z}
    \right) \right]^{-1} \equiv \\
    &\equiv \frac{\i \hbar}{2 \pi} \, \frac{1}{\hbar \omega -
    E_{\vec{k}}} \,. \nn
\end{flalign}
We obtain the energy spectrum in a bulk monomolecular crystal
[13,14] by calculating real part of the pole of this Green's
function:
\begin{equation}
    E_{\vec{k}} = \Delta + 2 \left( X_{x} \cos a_{x} k_{x} + X_{y}
    \cos a_{y} k_{y} + X_{z} \cos a_{z} k_{z} \right) \,.
\end{equation}
In order to perform comparison with dispersion law of
excitons in the film, this expression will be written in a
simpler ($X_{x} = X_{y} = X_{z} \equiv - |X|$, $a_{x} = a_{y} =
a_{z} \equiv a$) and non-dimensional form:
\begin{equation}
    \mathcal{E}_{\vec{k}} \equiv \frac{E_{\vec{k}} - \Delta}{|X|} =
    \mathcal{F}_{xy} + \mathcal{G}_{z} \,,
\end{equation}
where
\begin{equation*}
    \mathcal{F}_{xy} = - 2 \left( \cos a k_{x} + \cos a k_{y}
    \right) \,; \r \mathcal{G}_{z} = - 2 \cos a k_{z} \,.
\end{equation*}

This dispersion law is shown in Figure 1, as a function of
two-dimensional value $\mathcal{F}_{xy}$:
\begin{equation*}
    \mathcal{E}_{\vec{k}} = \mathcal{E}_{z} \left( \mathcal{F}_{xy} \right)
    \,,
\end{equation*}
It is clear that for $a k_i \in [0, + \pi]$, $i=x,y,z$ (the
first Brilouin zone), these values are within the intervals:
\begin{equation*}
    \left. \ba{r} \mathcal{F}_{xy} \in [-4, +4] \,; \vx \\
    \mathcal{G}_{z} \in [-2, +2] \,, \ea \, \right\} \, \Rightarrow \,
    \mathcal{E}_{\vec{k}} \in [-6, +6] \,.
\end{equation*}
The presence of permitted (continuous) energy levels is visible. \vy

\begin{figure}[h]
 \centering
 \epsffile{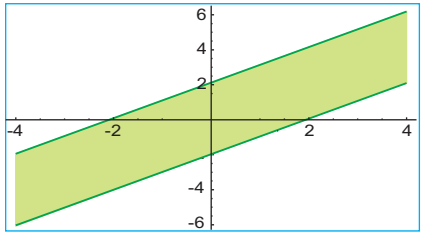}
 \caption{Exciton dispersion law of bulk crystal}
\end{figure}

Since molecular crystals are dielectric, it is essential to
determine relative permittivity of these structures. The dynamic
permittivity is defined by systems response to external
perturbation [1,15,20], by expression:
\begin{equation}
    \varepsilon^{-1} (\omega) = 1 - \frac{\pi S |X|}{\i \hbar} \,
    \left[ \, \Gamma (\omega) + \, \Gamma (- \omega) \right] \,,
\end{equation}
where $S$ is a frequency property of a given crystal and external
variable electromagnetic field. As it was mentioned before, in
this (zero) approximation Paulian Green's functions ($\Gamma$) are
transforming into Bosonian ones ($G$), therefore:
\begin{equation}
    \varepsilon^{-1} (\omega) = 1 - \frac{\pi S |X|}{\i \hbar} \,
    \left[ G (\omega) + G (- \omega) \right] \,.
\end{equation}
Substituting (22) with (23) in (26) and by denoting $\ds
\frac{\hbar \omega}{|X|} \equiv |f|$ and $\ds \frac{\Delta}{|X|}
\equiv |p|$, we obtain the expression for dynamic permittivity in
a bulk sample of monomolecular crystal (molecular crystal with simple cell):
\begin{equation}
    \r \varepsilon_{\vec{k}} (\omega) = \left[ 1 + S |X| \,
    \frac{E_{\vec{k}}} { (\hbar \omega)^{2} - E_{\vec{k}}^{2}}
    \right]^{-1} \equiv \left[ 1 + S \, \frac{\mathcal{E}_{\vec{k}} +
    |p|} { |f|^{2} - \left( \mathcal{E}_{\vec{k}} + |p| \right)^{2}}
    \right]^{-1} .
\end{equation}

Dependence of relative dynamic permittivity, expression (27), on
reduced frequency (non-dimensional factor $\hbar \omega /
\Delta$) of the external electromagnetic field is shown in Figure
2.


The presence of a single absorption zone is visible within certain
boundary frequencies. This energy zone was calculated for
two-dimension center of Brilouin's zone ($k_x=k_y=0$; \
$k_z=[0,\pi]$). For all other energies this crystal is transparent
and has no spatial non-homogeneousness. \vy

\begin{figure}[h]
 \centering
 \epsffile{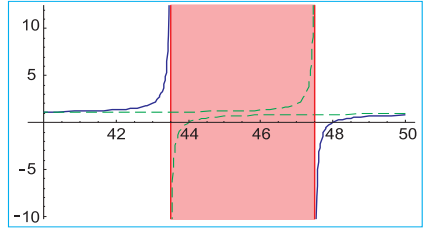}
 \caption{Relative permittivity of bulk crystal}
\end{figure}

\np

\section{Excitons in thin film-structures}

Opposite to ideal unlimited structures, real crystals have no
translation invariance property. The presence of certain boundary
conditions is one of reasons for symmetry breaking [4$-$7]. Lets
observe the ideal ultrathin film with simple cubic structure, made
in the substrate i.e. by doping process. Here, by the term "ideal",
we wanted to denote that there exist no breaking of inner crystal
structure (no defects, ingredients, etc), and not as in no spatial limits.
The film dimensions are such that it is unlimited in $XY$ planes, and in
z-axis has final thickness ($L$). This means that this film has two
unlimited boundary planes parallel to $XY$ planes, for $z = 0$ and
$z = L \equiv n_z$. \vy

\subsection{The model}

The film-structure with primitive crystalline lattice (one molecule
per elementary cell): mono\-molecular crystalline film, with
specified parameters is shown in Figure 3.

Since boundary planes of the film are taken as being normal to
$z$-axis, the index of parallel $XY$ planes $n_z$ has values $n_{z}
= 0, 1, 2, \, \ldots \,, \, N$ where $N \in [ 2, \, 8 ]$ is for ultra
thin films. Indices $n_x$ and $n_y$, determining position of a
molecule in every $XY$ plane may have arbitrary whole-number values
(practically from $- \infty$ to $+ \infty$).


\begin{figure}[h]
 \centering
 \epsffile{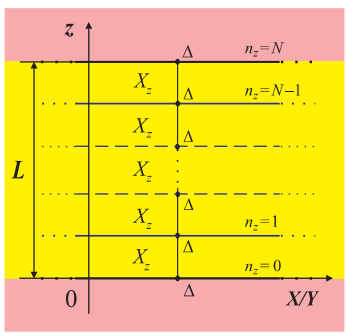} \vl\vm
 \caption{Cross-section of crystalline film-model}
\end{figure}

For calculating exciton energies in this film, we start from
equation (19) on which, due to spatial limitations of the film at
$z$-direction, we may apply partial spatial Fourier transformation:
\begin{equation} \ba{c} \ds
    \delta_{\vec{n} \vec{m}} = \frac{1}{N_{x}N_{y}} \sum_{k_{x}
    k_{y}} \e^{ik_{x}a_{x} (n_{x}-m_{x})}
    \e^{ik_{y}a_{y}(n_{y}-m_{y})} \delta_{n_{z}m_{z}}
    \vx \\ \ds
    f_{\vec{n} \vec{m}}(\omega) = \frac{1}{N_{x} N_{y}}
    \sum_{k_{x} k_{y}} \e^{ik_{x}a_{x} (n_{x}-m_{x})}
    \e^{ik_{y}a_{y} (n_{y}-m_{y})} f_{n_{z}m_{z}} (k_{x},k_{y},
    \omega) \ea
\end{equation}
(only along $x$ and $y$ directions). For expression shortening , it
is convenient to introduce markings
$G_{n_{z}m_{z}}(k_{x},k_{y},\omega) \equiv G_{n_{z}m_{z}}$. For $\ds
X_{x}=X_{y}=X_{z} \equiv - \left| X \right|$, $\ds a_{x}=a_{y}=a_{z}
\equiv a$ and $\ds \mathcal{K} = \frac{i \hbar}{2 \pi \left| X
\right|} \delta_{n_{z}, m_{z}}$, we obtain:
\begin{equation}
    G_{n_{z}-1,m_{z}} + \varrho \, G_{n_{z}m_{z}} + G_{n_{z}+1,m_{z}}
    = \mathcal{K} \delta_{n_{z}, m_{z}} \,,
\end{equation}
where denotation is introduced:
\begin{flalign}
    \varrho &= \frac{\hbar \omega - \Delta}{\left| X \right|} + 2
    \left( \cos a k_{x} + \cos a k_{y} \right) \equiv \nn \\
    &\equiv \frac{E_{\vec{k}} - \Delta}{\left| X \right|} -
    \mathcal{F}_{xy} \equiv \mathcal{E}_{\vec{k}} - \mathcal{F}_{xy}
\end{flalign}
(the quantity $\mathcal{F}_{xy}$ is defined in expression (24),
while $\mathcal{E}_{\vec{k}}$ i.e. $E_{\vec{k}}$ which express a
possible energy of excitons in films, will be calculated latter).

Equation (29) is in fact a system of $N+1$ nonhomogeneous
algebraic differential equations with (starting boundary)
conditions: $G_{n_{z}, m_{z}} = 0$, for $n_{z} < 0$ and $n_{z} >
N+1$:
\begin{flalign}
    \varrho \, G_{0} + G_{1} &= \mathcal{K}_{0} \nn \\
    G_{0} + \varrho \, G_{1} + G_{2} &= \mathcal{K}_{1} \nn \\
    G_{1} + \varrho \, G_{2} + G_{3} &= \mathcal{K}_{2} \nn \\
    \cdots \;\r \cdots \;\r \cdots \; & \;\r \cdots \nn \\
    G_{n_{z}-2} + \varrho \, G_{n_{z}-1} + G_{n_{z}}
    &= \mathcal{K}_{n_{z}-1} \nn \\
    G_{n_{z}-1} + \varrho \, G_{n_{z}} + G_{n_{z}+1}
    &= \mathcal{K}_{n_{z}} \\
    G_{n_{z}} + \varrho \, G_{n_{z}+1} + G_{n_{z}+2}
    &= \mathcal{K}_{n_{z}+1} \nn \\
    \cdots \r\r\r \cdots \r\r\r \cdots \r & \r\r \cdots \nn \\
    G_{N-3} + \varrho \, G_{N-2} + G_{N-1}
    &= \mathcal{K}_{N-2} \nn \\
    G_{N-2} + \varrho \, G_{N-1} + G_{N}
    &= \mathcal{K}_{N-1} \nn \\
    G_{N-1} + \varrho \, G_{N} &= \mathcal{K}_{N} \nn
\end{flalign}
where is describe: $G_{n_{z}} \equiv G_{n_{z}m_{z}}$ and
$\mathcal{K} \delta_{n_{z}, m_{z}} \equiv \mathcal{K}_{n_{z}}$,
because $m_{z}$ is blinding index.

\subsection{Dispersion law}

In order to find exciton energies, we need poles of Green's
functions, which are obtained when a determinant of a system (31)
is equalized with zero, i.e.
\begin{equation}
    D_{N+1} (\varrho) = \left| \ba{ccccccccc} \varrho & 1 & 0
    & 0 & \; \cdots \; & 0 & 0 & 0 & 0 \\
    1 & \varrho & 1 & 0 & \cdots & 0 & 0 & 0 & 0 \\
    0 & 1 & \varrho & 1 & \cdots & 0 & 0 & 0 & 0 \\
    & & & & & & & & \\
    \cdot & \cdot & \cdot & \cdot & \ddots & \cdot & \cdot & \cdot & \cdot \\
    & & & & & & & & \\
    0 & 0 & 0 & 0 & \cdots & 1 & \varrho & 1 & 0 \\
    0 & 0 & 0 & 0 & \cdots & 0 & 1 & \varrho & 1 \\
    0 & 0 & 0 & 0 & \; \cdots \; & 0 & 0 & 1 & \varrho \ea
    \right|_{N+1} \equiv 0 \,,
\end{equation}
and this determinant is actually a note for Chebyshev's polynomials of second
kind (and ($N+1$)-th order) [21,22]:
\begin{equation}
    D_{N+1} (\varrho) \equiv \mathcal{C}_{N+1} (\varrho) = \varrho \,
    \mathcal{C}_{N} (\varrho) - \mathcal{C}_{N-1} (\varrho) \,.
\end{equation}
The condition $D_{N+1} (\varrho) = 0$ is reduced to
$\mathcal{C}_{N+1} (\varrho) = 0$ and is satisfied by $N+1$
solutions in form:
\begin{equation}
    \varrho_\nu = - 2 \, \cos \frac{\nu \, \pi}{N+2} \,; \r \nu =
    1,2, \ldots , N+1 \,.
\end{equation}
Using this and replacing equation (30) we find:
\begin{equation}
    E_{k_x k_y}(\nu) = \Delta - 2 |X| \left( \cos a k_{x} + \cos a
    k_{y} + \cos \frac{\pi \nu}{N +2} \right) \,.
\end{equation}

In order to compare with dispersion law of excitons in a bulk we
will write this expression in more simple, non-dimensional form
($\varrho_{\nu} = - 2 \cos a_z k_z (\mu)$):
\begin{equation}
    \mathcal{E}_{k_{x}k_{y}}(\nu) = \mathcal{F}_{xy} + \mathcal{G}_{z}(\nu)
    \,; \r \mathcal{G}_{z} (\nu) \equiv \frac{1}{2} \left[1 -
    \cos a_{z} k_{z} (\nu) \right] \,.
\end{equation}

Previous expression represents the dispersion law of excitons of
ideal monomolecular film and has the same form as the expression
(24) obtained for corresponding ideal unlimited structures, with
difference that in (24) $k_z$ is practically continuous
variable (interval $[0, \pi/a]$) as $k_{x} $ and $k_{y}$, while here
is discrete and is given by expression:
\begin{equation}
    k_{z}(\nu) = \frac{\pi}{a} \frac{\nu}{N +2} \,; \r \nu = 1,2,
    \, \ldots \,, N+1 \,.
\end{equation}

Graphical representation of dispersion law is given on Figure 4, showing
possible exciton energies in ideal five-layer monomolecular film
(full lines) together with bulk boundaries (dotted lines). Same as for
corresponding bulk crystalline structures, we will use ordinate for
values of reduced non-dimensional energies $\ds \mathcal{E}_{\nu}$,
depending on two-dimensional function $\mathcal{F}_{xy}$, for which
is used graph abscissa.

\begin{figure}[h]
 \centering
 \epsffile{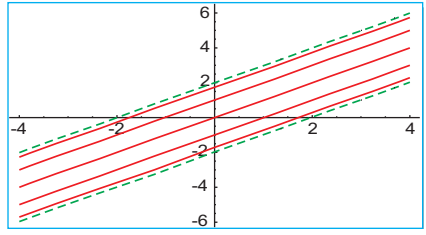} \vn
 \caption{Exciton energies of molecular film}
\end{figure}

These analyzes have shown significant differences regarding dispersion law for
excitons in spatially strongly limited systems (nanofilm-structures)
as strictly result of border presence of these structures, in which
energy spectra is highly discrete and have two gaps. Sizes of
gaps depends on film thickness and decrease rapidly with its
increase.

\subsection{Spectral weights}

In order to find certain Green's function we will start from the
system of equations (29), which is now suitable to represent in the
operator form:
\begin{equation}
    \hat{\mathcal{D}}_{N+1} \hat{\mathcal{G}}_{N+1} = \hat{
    \mathcal{K}}_{N+1} \;\; \Longrightarrow \;\; \hat{\mathcal{G}}_{N+1} =
    \hat{\mathcal{D}}_{N+1}^{-1} \hat{\mathcal{K}}_{N+1} \,,
\end{equation}
where: $\hat{\mathcal{D}}_{N+1}$ is a matrix corresponding to
the determinant of system $D_{N+1}$, $\hat{\mathcal{G}}_{N+1}$ and $
\hat{\mathcal{K}}_{N+1}$ are vectors of Green's functions and
Kronecker's delta symbols:
\begin{equation} \ba{r} \ds
    \tilde{\mathcal{G}}_{N+1} = \left( \begin{array}{c}
                                         G_{0,m_z} \\
                                         G_{1,m_z} \\
                                           \; . \; \\
                                           \; . \; \\
                                           \; . \; \\
                                         G_{n_z,m_z}
                                           \; . \; \\
                                           \; . \; \\
                                           \; . \; \\
                                         G_{N,m_z}
                              \end{array} \right) ; \r
    \ds \tilde{\mathcal{K}}_{N+1} = -\frac{i \hbar}{2 \pi \left| X
    \right|} \left( \begin{array}{c}
                                         \delta_{0,m_z} \\
                                         \delta_{1,m_z} \\
                                           \; . \; \\
                                           \; . \; \\
                                           \; . \; \\
                                         \delta_{n_z,m_z}
                                           \; . \; \\
                                           \; . \; \\
                                           \; . \; \\
                                         \delta_{N,m_z}
                              \end{array} \right) . \ea
\end{equation}
Since the inverse matrix $\hat{\mathcal{D}}_{N+1}^{-1}$ may be
expressed through adjunct matrix, whose elements $D_{ik}$ are
cofactors of elements $d_{ik}$ from the direct matrix, we may write:
\begin{flalign}
    G_{n_z,m_z} &= \frac{1}{\mathcal{D}_{N+1}} \sum_q D_{n_z,q}
    K_{q,m_z} = \nn \\
    &= - \frac{1}{\mathcal{D}_{N+1}} \frac{i \hbar}{2 \pi
    \left| X \right|} \sum_q D_{n_z,q} \delta_{q,m_z} = \\
    &= - \frac{i \hbar}{2 \pi \left| X \right|}
    \frac{D_{n_z,m_z}}{\mathcal{D}_{N+1}} \,. \nn
\end{flalign}
Cofactor $D_{n_z,m_z}$ calculation is based on knowing
the system determinant $D_{N+1}$.

Since for the equilibrium processes within system are important only diagonal Green's functions $G_{n_z;n_z} \equiv G_{n_z}$, calculating
cofactors $D_{n_z,m_z} \equiv D_{n_z}$ is significantly
simplified. It turns out that they are equal to the product of two auxiliary determinants:
\begin{equation}
    D_{n_z} = \mathcal{C}_{n_z} \mathcal{C}_{N-n_z} \,,
\end{equation}
where $\mathcal{C}_{n_z-1}$ and $\mathcal{C}_{N-n_z}$ are
corresponding Chebyshev's polynomials of second kind. Therefore,
Green's function of the ideal film is:
\begin{equation}
    G_{n_z} = \frac{i \hbar}{2 \pi \left| X \right|}
    \frac{\mathcal{C}_{n_z}\mathcal{C}_{N-n_z}}{\mathcal{C}_{N+1}} \,.
\end{equation}

These Green's functions are multipolar, since denominator consists of a polynomial $\mathcal{C}_{N+1}$ of $N+1$-th order. Therefore
factorization on simple poles must be performed [21]:
\begin{equation}
    G_{n_z} = - \frac{i \hbar}{2 \pi |X|} \sum_{\nu=1}^{N+1}
    \frac{g_{n_z;n_z}(\varrho_{\nu})}{\varrho - \varrho_{\nu}} \,.
\end{equation}
Spectral weights $g_{n_z;n_z}(\varrho_{\nu}) \equiv
g_{n_z}^{\nu}$ then may be expressed using:
\begin{equation}
    g_{n_z}^{\nu} = \frac{\mathcal{C}_{n_z}(\varrho_{\nu})
    \mathcal{C}_{N-n_z}(\varrho_{\nu})} {\ds \left. \frac{d}{d
    \varrho}\mathcal{C}_{N+1}(\varrho) \right|_{\varrho =
    \varrho(\nu)}} \,.
\end{equation}
Using the rule for derivative of determinant we obtain:
\begin{flalign}
    & \left[ \frac{d}{d \varrho} \mathcal{C}_{N+1} (\varrho)
    \right]_{\varrho = \varrho_{\nu}} = \sum_{i=1}^{N+1}
    \mathcal{C}_{i-1}(\varrho_{\nu}) \mathcal{C}_{N+1-i}(\varrho_{\nu})
    = \nn \\
    &= \sum_{i=1}^{N+1} \frac{\sin i \xi_{\nu} \sin (N+2-i)
    \xi_{\nu}}{\sin^2 \xi_{\nu}} = \\
    &= \frac{\sin^2 (n_z +1) \xi_{\nu}}{\ds \sum\dl_{i=1}^{N+1}
    \sin^2 i \xi_{\nu}} \,; \r \xi_{\nu} = \frac{\pi \nu}{N+2}
    \,, \nn
\end{flalign}
and spectral weights became:
\begin{equation}
    g_{n_z}^{\nu} = \frac{2}{N+2} \sin^2 \left[ (n_z+1 ) \frac{\pi
    \nu}{N+2} \right] \,.
\end{equation}

The spectral weights of Green's functions are squares of the
modules of the wave function of excitons [1$-$3,13$-$15] and enable
determination of spatial distribution, i.e. probability to find
excitons with certain energies per layers of crystalline film. This
is in fact the spatial distribution of probability to find certain
energy state of excitons.

Numerically calculated, values of reduced energies and
corresponding spectrum functions (spatial distribution of
probability) for four-layered film ($N=4$, where $k_x = k_y = 0$)
are shown in the table.

Table 1. shows spatial distribution of exciton
energies occurrence probabilities in ideal monomolecular film.

\fns

\begin{table}[h]
  \centering
\begin{tabular}{||r||c|c|c|c|c||} \hline \hline
\fns \bf Reduced \ \
& \multicolumn{5}{|c||}{\fns \bf ULTRATHIN FILM} \vn \\
\fns \bf relative \hy
& \multicolumn{5}{|c||}{\fns \bf a t o m i c \ \ p l a n e} \\
\cline{2-6} \fns \bf \ ENERGY \ &
    $\ds \ba{c} \mbox{\fns upper} \vn \\ \mbox{\fns boundary} \ea$ &
    $\ds \ba{c} \mbox{\fns first} \vn \\ \mbox{\fns inner} \ea$ &
    $\ds \ba{c} \mbox{\fns central} \vn \\ \mbox{\fns (inner)} \ea$ &
    $\ds \ba{c} \mbox{\fns last} \vn \\ \mbox{\fns inner} \ea$ &
    $\ds \ba{c} \mbox{\fns lower} \vn \\ \mbox{\fns boundary} \ea$ \\
\hline\hline $-$1,73205 \hy & 0,08333 & \hy 0,25000 \hy
& \hy 0,33333 \hy & \hy 0,25000 \hy & 0,08333 \\
\hline $-$1,00000 \hy & 0,25000 & \hy 0,25000 \hy & \hy
0,00000 \hy & \hy 0,25000 \hy & 0,25000 \\
\hline 0,00000 \hy & 0,33333 & \hy 0,00000 \hy & \hy
0,33333 \hy & \hy 0,00000 \hy & 0,33333 \\
\hline 1,00000 \hy & 0,25000 & \hy 0,25000 \hy & \hy
0,00000 \hy & \hy 0,25000 \hy & 0,25000 \\
\hline 1,73205 \hy & 0,08333 & \hy 0,25000 \hy & \hy
0,33333 \hy & \hy 0,25000 \hy & 0,08333 \\
\hline\hline
    \end{tabular} \vy
    \caption{\nrs Exciton probabilities in the ideal four-layered film}
\end{table}

\nrs This table shows that for one certain energy, probability of
exciton occurrence per all layers is equal to one, and that
probability per one layer for all energies is also equal to one,
i.e.
\begin{equation}
    \sum_{n_z =0}^{N} g_{n_z}^{\nu} = 1 \,; \r \sum_{\nu =1}^{N+1}
    g_{n_z}^{\nu} = 1 \,.
\end{equation}

\subsection{The permittivity}

While determining dynamic permittivity of crystalline film, the formula
by Dzyaloshinski and Pitaevski [20] may also be used in the same
form (26) in which was used for calculating permittivity of
corresponding bulk structures, with difference that in this case
permittivity depends on a film layer $n_z$, i.e:
\begin{equation}
    \varepsilon_{n_z}^{-1} (\omega) = 1 - \frac{\pi \, S \,
    |X|}{\i \,     \hbar} \, \left[ G_{n_z} (\omega) + G_{n_z}
    (- \omega) \right] \,.
\end{equation}
Since starting Hamiltonian was taken in harmonic approximation,
ignoring (small) members of exciton-phonon interaction [1,6,15],
permittivity tensor has the real elements only. It is led to that
all elements of permittivity tensor in one crystalline plane parallel
to boundary planes are mutually equivalent, i.e. they depend on
plane position ($n_z \in [0,N]$).

Substituting expression for Green's functions (43) to (46), we obtain
expression for the relative dynamic permittivity tensor elements in
direction normal to boundary planes in form:
\begin{equation}
    \varepsilon_{n_z}^{-1} = 1 - \frac{S}{2} \sum_{\nu=1}^{N+1}
    \sum_{s = +,-} \frac{g_{n_z}^{\nu}}{\varrho_{s} -
    \varrho_{\nu}} \,,
\end{equation}
where: $\ds \varrho_{\pm} = \mp |f| - |p| - \mathcal{F}_{xy}$, and
after arranging the expression we finally follow:
\begin{equation}
    \varepsilon_{n_z}(\omega)= \left[ 1 + S \, \sum_{\nu=1}^{N+1}
    g^{\nu}_{n_z} \frac{ \varrho_{\nu} - |p| +
    \mathcal{F}_{xy}}{|f|^2 - \left( \varrho_{\nu} - |p| + \mathcal{F}_{xy}
    \right)^2} \right]^{-1} \,.
\end{equation}

Figures 5 a-c show a dependence of dynamical permittivity
($\varepsilon$) on reduced relative energy, i.e. the frequency of
external electromagnetic field ($f \equiv \hbar \omega / |X|$) for
four-layered monomolecular film. Dependence was calculated for plane
center of the Brilouin's zone ($k_x=k_y=0$), but individually per
atom planes (parallel boundary areas) of crystalline film, that is
for $n_z = 0$, and $n_z = 1,2,3,4$.
Each graph shows number and position of resonating peaks. The
resonating peaks on frequency dependence of dynamic permittivity are
the positions $-$ resonating frequencies where permittivity diverges
to $\pm \infty$. These are also the energies (wavelengths) of
electromagnetic radiation which model crystal in given place
practically "swallows", i.e. these energies are absolutely absorbed
there.

\begin{figure}[hbt]
 \centering
 \epsffile{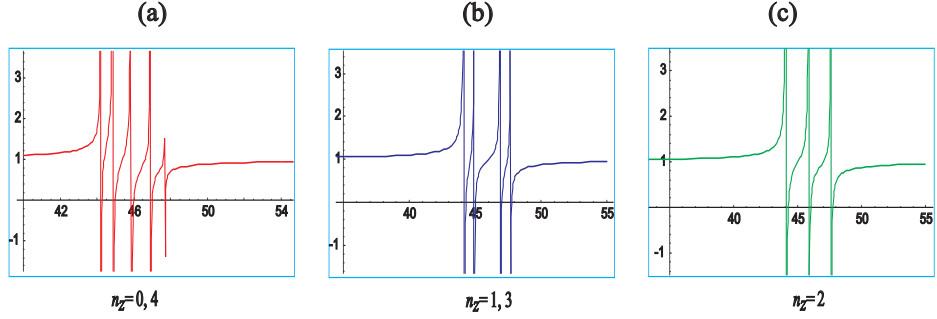}
 \caption{Dynamic permittivity of ideal films} \vx
\end{figure}

A final number of resonating peaks is present, because $k_z$
component of exciton wave vector is discrete. Only for certain
values of $k_z$ component for wave vector (with given values $k_x$
and $k_y$), resonating and radiation absorption may occur.
Therefore, at least three and maximum five peaks occur - number is
equal to number of permitted states along axis of translation
symmetry breakage, in this case along z-axis. Due to the symmetry of
model, distribution of peaks in complementary planes is the same:
$n_z = 0$ and $n_z = 4$ (two boundary planes), as well as for $n_z =
1$ and $n_z = 3$ (first two inner planes). Only $n_z = 2$ is
different, since for the odd total number of atomic planes the
middle plane has no complementary plane!

This result may be explained by experimental facts regarding
resonating optical peaks in similar molecular layered
nanostructures. In papers [23$-$25] this was evidenced in perylene
chemical compounds (PTCDA, PTCS, PTFE and PBI) and explained by
resonating effects at specific unoccupied levels. These effects are
manifested by narrow optic absorption in close infra red band. In
comparison with our results, which are concerning deeper infra red band
in electromagnetic radiation, it may be concluded that these
differences are effect of differences in crystalline (chemical and
physical) structure as well as in the model (real and ideal) of
samples investigated.

\np

\section{Conclusion}

This paper describes original application of Paulian Green's
function method onto theoretical studies of optical properties of molecular crystals.
Using Boson representation, microscopic (dispersion law and exciton state distribution)
and macroscopic (dynamic permittivity) properties of these crystals may be
successfully described.

Strictly following defined procedure, this paper shows that this
method may be adapted and successfully applied to the study of
dielectric properties in structures with disturbed
spatial-translational symmetry, such as ultra thin films.
Functioning of newly developed approach is illustrated through the problem
of finding energy spectra and exciton states, as well as for determining relative
permittivity of ideal monomolecular film.

In presence of two parallel boundaries in the system, energy
spectrum is determined and possible exciton states were found.
Important differences, comparing with unlimited crystalline
structures, have been observed. Energy spectrum of excitons in
monomolecular films is explicitly discrete, and the number of
discrete levels is equal to the number of atomic planes (including
boundary areas) along the axis of the spatial limitations of ultra
thin film. In bulk sample, a single zone where excitons "take"
all possible energy values exists. All discrete levels are located
within bulk boundaries, and the difference in zone width
depends strictly (and inversely) on film thickness.

Comparing with bulk structures, where excitons may be found at
any place with equal probability, in monomolecular film
structures probability of finding exciton strongly depends on
film thickness.

In exciton systems of monomolecular crystal bulk, where relative
dynamic permittivity depends on frequency, continuous absorption
zone exists in certain range of external radiation energy. In
monomolecular film-structures resonating peaks exist with
precisely determined energies, i.e. resonating frequencies.
Number of these peaks depends on position of atomic plane
(regarding boundary planes of the film) for which permittivity is
being calculated: it decreases with the depth of ultra thin
film.

Differences between properties of observed film and corresponding
bulk structures drastically decrease as the thickness of film is higher.
All this implies to the existence, and is a consequence, of the quantum size
effects.

Method of Green's differential functions, adapted on described manner should be
applied further to the study of behavior and properties of more realistic
quantum structures, for instance ultra thin films with perturbed
boundary conditions.

\subsection*{Acknowledgements}

Investigations whose results are presented in this paper were
partially supported by the Serbian Ministry of Sciences (Grant No
141044) and by the Ministry of Sciences of the Republic of Srpska.


\bibliographystyle{my-h-elsevier}

\end{document}